# PEERS: A Parallel and Exact Effective Resistance Solver via Implicit Inversion and Augmented Symbolic Analysis

Baiyu Chen, Lin Gan, Guangwen Yang*, and Wenjian Yu*, *Fellow, IEEE*

***Abstract*—High-precision effective resistance computation is a cornerstone of Electronic Design Automation (EDA) sign-off, yet it remains a fundamental bottleneck in large-scale power grid analysis, spectral sparsification, and circuit reliability. Existing approaches face a prohibitive "precision-memory impasse": approximate methods lack the stringent accuracy required for high-stakes industrial sign-off, while exact methods either suffer from redundant query overheads or trigger $\mathcal{O}(n^2)$ memory explosions. To resolve this, we propose PEERS, a Parallel and Exact Effective Resistance Solver powered by an implicit inverse computing model of the Cholesky factor. By integrating a state-inherited augmented depth-first search (DFS) with a dynamic query update mechanism, PEERS eliminates numerical redundancy and evaluates all-edge resistance queries in a single parallel sweep. We provide a rigorous Work-Span analysis, proving that for graphs satisfying an $\mathcal{O}(n^\alpha)$ separator theorem, PEERS achieves a theoretically optimal parallel span of $\mathcal{O}(n^\alpha)$ while strictly maintaining $\mathcal{O}(nnz(L))$ space complexity. Numerical evaluations on industrial benchmarks demonstrate that PEERS achieves an average speedup of 83.3× over state-of-the-art parallel solvers under identical memory constraints. Notably, PEERS processes a 1-million-node industrial graph in just 18.8 seconds and scales to 17 million nodes in under an hour, providing the first computationally feasible path for exact all-edge resistance analysis in multi-million-gate designs.**

## I. Introduction

Effective resistance serves as a cornerstone metric in spectral graph theory and circuit analysis. For a weighted graph $G = (V, E, W)$ interpreted as a resistive network, the effective resistance $R_{uv}$ between nodes $u$ and $v$ corresponds to the potential difference induced by a unit current injection-extraction pair. Beyond its physical interpretation in Kirchhoff's laws, effective resistance has emerged as a vital analytical tool for characterizing graph connectivity, structural robustness, and spectral invariants.

The seminal work of Spielman and Srivastava [1] demonstrated that all edge effective resistances are the fundamental weights required to construct spectral sparsifiers that preserve the Laplacian quadratic form. Consequently, this metric has found extensive application across the Electronic Design Automation (EDA) landscape, including power-grid IR-drop analysis [2], graph clustering [3], algebraic multigrid preconditioning, and recently, mitigating over-squashing in graph neural networks (GNNs) [4]. In these contexts—particularly for industrial power delivery networks (PDNs)—high numerical precision is non-negotiable, as approximation artifacts can lead to pessimistic timing violations or catastrophic reliability oversights.

Despite its utility, the exact evaluation of all-edge effective resistances remains a formidable computational bottleneck for large-scale systems. Given a graph Laplacian $L_G$, the resistance between nodes $u$ and $v$ is typically computed as $R_{uv} = (e_u - e_v)^T L_G^\dagger (e_u - e_v)$. For a query set of $m$ edges, a naive exact approach necessitates $m$ independent sparse linear system solves. For modern industrial-scale graphs containing millions of nodes ($n$) and edges ($m$), this $\mathcal{O}(m)$ solve-dependency is computationally prohibitive.

Existing strategies to mitigate this cost generally fall into two categories, each with inherent limitations:

- **Approximate Methods:** Techniques such as random sketching [1], [5] and random walks [6], [7] reduce work by sacrificing exactness. However, their complexity scales poorly with the target error tolerance $\epsilon$ ($\mathcal{O}(\epsilon^{-2})$), and they lack the strict numerical reliability required for "sign-off" quality circuit analysis.
- **Direct and Inverse-based Methods:** Sparse direct solvers leverage Cholesky factorization to provide exact solutions. While reliable, repeated triangular solves incur massive redundant traversals of the factor structure. Conversely, explicit inverse-factorization methods [8] attempt to precompute $L^{-1}$, but they trigger an $O(n^2)$ memory explosion as the inverse Cholesky factor loses the sparsity of the original factor.

Parallelization offers a potential remedy, yet it is often hampered by the irregular memory access patterns of sparse traversals. Domain decomposition [9] can introduce dense Schur complements that negate parallel gains, while standard task-based parallelization of $m$ queries often suffers from severe thread-contention and synchronization overheads.

To break this precision-memory impasse, we propose **PEERS** (Parallel and Exact Effective Resistance Solver). PEERS is founded on the insight that the numerical entries of the inverse Cholesky factor can be computed *implicitly* by traversing the symbolic sparsity pattern of the original factor. Unlike prior work, PEERS avoids the $O(n^2)$ storage of $L^{-1}$ by using a dynamic accumulation mechanism that evaluates resistance contributions in a single parallel sweep.

The authors are with the Department of Computer Science and Technology, Tsinghua University, Beijing 100084, China (e-mail: chenby23@mails.tsinghua.edu.cn; lingan@tsinghua.edu.cn; ygw@tsinghua.edu.cn; yu-wj@tsinghua.edu.cn).

*Corresponding authors.

TABLE I
COMPLEXITY AND ACCURACY COMPARISON OF EFFECTIVE RESISTANCE SOLVERS FOR ALL-EDGE QUERIES

| Method Category | Representative | Factorization | All-Edge Query Work | Parallel Span | Space | Accuracy |
|---|---|---|---|---|---|---|
| Approximate | Random Walks | N/A | $\mathcal{O}(m \log^c n/\epsilon^2)$ | High [1] | $\mathcal{O}(m)$ | $\epsilon$-approx. |
| Equation-Based | LSolve [10] [2] | $\mathcal{O}(W_{chol})$ | $\mathcal{O}(mn + \sum_{e\in E} nnz(L_{\text{reach}}(e)))$ | $\mathcal{O}(n^\alpha)$ [2] | $\mathcal{O}(nnz(L) + P \cdot n)$ [2] | Exact |
| Explicit Inverse | INV [8] [3] | $\mathcal{O}(n^3)$ | $\mathcal{O}(mn)$ | $\mathcal{O}(n)$ [3] | $\mathcal{O}(n^2 + Pn)$ | Exact |
| **Implicit** | **PEERS** | $\mathcal{O}(\mathbf{W_{chol}})$ | $\mathcal{O}(\mathbf{n^\alpha nnz(L)})$ | $\mathcal{O}(\mathbf{n^\alpha})$ | $\mathcal{O}(\mathbf{nnz(L) + P \cdot n^\alpha})$ | **Exact** |

[1] This is due to irregular memory access patterns and high synchronization overhead.
[2] Under the theoretical Work-Span model ($P \to \infty$), LSolve achieves $\mathcal{O}(n^\alpha)$ span. However, initializing dense right-hand side vectors per query incurs an $\mathcal{O}(mn)$ work penalty. Furthermore, achieving optimal span triggers a memory explosion because each of the $P$ concurrent threads requires an independent $\mathcal{O}(n)$ workspace. If space is strictly bounded to $\mathcal{O}(nnz(L))$, parallel execution is severely bottlenecked into sequential batches.
[3] Explicit inverse formulation recursively computes rows from $n$ down to 1. Each row strictly depends on subsequent rows, creating a linear dependency chain that limits the span to $\mathcal{O}(n)$. Additionally, explicit inversion introduces severe dense fill-in, driving computational work to $\mathcal{O}(n^3)$ and destroying memory sparsity.
* *Notation:* $W_{chol}$ denotes the work complexity of sparse Cholesky factorization. $P$ denotes the number of parallel threads. $\alpha$ is the graph separator exponent (e.g., $\alpha = 1/2$ for planar graphs and $\alpha = 2/3$ for 3D grids).

As summarized in Table I, PEERS breaks this precision-memory impasse by avoiding the $\mathcal{O}(n^2)$ storage of $L^{-1}$, yielding exact resistance values while matching the shallow $\mathcal{O}(n^\alpha)$ parallel span of nested dissection. Experimental evaluations on industrial benchmarks with up to 17 million nodes demonstrate that PEERS reshapes the scalability of exact solvers. On a 1.6-million-edge graph, PEERS completes all-edge queries in 18.8 seconds—an 83.3× speedup over state-of-the-art parallel solvers—while maintaining a memory footprint strictly bounded by the density of the Cholesky factor. In fact, the necessity for exact effective resistance in PEERS extends beyond theoretical graph sparsification into critical VLSI sign-off tasks. In modern Power Delivery Network (PDN) analysis, effective resistance is the primary metric for identifying *weak links*—local areas with high resistive paths to the power source that are prone to IR-drop violations. Furthermore, Electromigration (EM) assessment requires high-precision current density modeling; even minor inaccuracies from approximate solvers can lead to optimistic life-time estimations, potentially resulting in post-silicon failures. PEERS provides the numerical "ground truth" required for these high-stakes reliability assessments at a scale previously reserved for approximate methods.

The primary contributions of this work are summarized as follows:

- **Implicit Inversion Architecture:** We develop a mathematical framework to compute exact resistances via the implicit inverse of the Cholesky factor, bypassing the $O(n^2)$ memory barrier.
- **Augmented Symbolic Analysis:** We introduce a reusable-state traversal mechanism that minimizes symbolic overhead by exploiting the transitive closure properties of the elimination tree.
- **Work-Span Theoretical Guarantees:** We provide rigorous proofs showing that for graphs with $O(n^\alpha)$ separators, PEERS achieves a near-optimal parallel span of $O(n^\alpha)$, ensuring scalability on massively parallel architectures.
- **Scalable EDA Benchmarking:** We demonstrate that PEERS can solve previously intractable industrial problems, providing a robust, zero-variance alternative to approximate solvers for large-scale circuit analysis.

## II. PRELIMINARIES AND RELATED WORK

In this section, we formalize the mathematical foundation of effective resistances and review the current state-of-the-art exact computation techniques. We specifically emphasize the structural relationship between the Laplacian matrix and the Cholesky factor, which underpins the efficiency of the proposed PEERS solver.

### A. Mathematical Foundations of Effective Resistances

Consider a weighted undirected graph $G = (V, E, w)$ with $|V| = n$ and $|E| = m$. The combinatorial Laplacian matrix $L_G \in \mathbb{R}^{n\times n}$ is defined as:

$$L_G = D - W = \sum_{(i,j)\in E} w_{i,j} e_{i,j} e_{i,j}^T, \quad (1)$$

where $D$ is the diagonal degree matrix, $W$ is the adjacency matrix, and $e_{i,j} = \mathbf{1}_i - \mathbf{1}_j$ is the incidence vector. By grounding an arbitrary node $g \in V$ (setting its potential to zero) and removing its corresponding row and column, $L_G$ becomes a reduced Laplacian matrix, which is Symmetric Positive Definite (SPD).

The effective resistance $R_{st}$ between nodes $s$ and $t$ is physically interpreted as the voltage drop induced between them by a unit current injection. Numerically, this is expressed using the inverse of the reduced Laplacian:

$$R_{st} = e_{s,t}^T L_G^{-1} e_{s,t}. \quad (2)$$

By performing a Cholesky factorization $L_G = LL^T$, where $L$ is a lower triangular matrix and $Z$ is the inverse of $L$, the effective resistance can be computed via two triangular substitutions ($z_{s,t} = Z_{:,s} - Z_{:,t}$):

$$R_{st} = e_{s,t}^T (LL^T)^{-1} e_{s,t} = \|L^{-1} e_{s,t}\|_2^2 = \|z_{s,t}\|_2^2. \quad (3)$$

Expanding this for the grounded cases yields the piecewise form:

$$R_{st} = \begin{cases} \|z_{s,t}\|_2^2, & s \neq g, t \neq g \\ \|z_s\|_2^2, & t = g \\ \|z_t\|_2^2, & s = g \end{cases} \quad (4)$$

### *B. Exact Computation and the Sparsification Bottleneck*

The exact computation of $\{R_e \mid e \in E\}$ is the computational cornerstone of spectral sparsification (Algorithm 1). As proved by Spielman and Srivastava [1], sampling edges with probabilities proportional to $w_e R_e$ yields a sparsifier that preserves the graph's spectral properties with high probability.

**Algorithm 1** Graph Sparsification by Effective Resistances

**Input:** $\mathcal{G} = (V, E, w)$: a weighted graph, $q$: expected number of added edges in the sparsifier.
**Output:** $\mathcal{S}$: a sparsifier.

1: Compute effective resistance $R_e$ of every edge $e \in E$ in graph $G$.
2: **for** $i = 1, 2, \cdots, q$ **do**
3: Choose a random $e$ of $\mathcal{G}$ with probability $p_e$ proportional to $R_e w_e$.
4: Add the chosen $e$ to $\mathcal{S}$ with weight $\frac{w_e}{q p_e}$.
5: **end for**
6: **Return** $\mathcal{S}$.

Despite its theoretical elegance, Step 1 remains the primary bottleneck for large-scale applications in VLSI and social networks. Current exact methods generally follow two paths:

**Algorithm 2** LSolve Algorithm [10]

**Input:** $L$: the Cholesky factor, $b$: the right-hand side, $G'$: the constructed graph.
**Output:** $x$: solution of $Lx = b$.

1: $\chi \leftarrow$ the set of nodes that can be reached in DFS from the nonzeros in $b$.
2: $x \leftarrow b$
3: **for** $j \in \chi$ **do**
4: **for** $i = j+1, j+2, \cdots, n$ and $L_{i,j} \neq 0$ **do**
5: $x_i \leftarrow x_i - L_{i,j} x_j$
6: **end for**
7: **end for**
8: **Return** $x$.

1) **Sparse Substitution (LSolve:** Since $e_{s,t}$ is a 2-sparse vector, the term $x = L^{-1} e_{s,t}$ can be solved using the LSolve algorithm [10]. LSolve utilizes the reachability in the factor graph to only visit nodes $k$ such that $(L^{-1}b)_k \neq 0$, yielding a complexity of $\mathcal{O}(nnz(L_{reach}))$. However, performing this for all $m$ edges results in a total work of $\mathcal{O}(m \cdot nnz(L_{reach}))$.
2) **Explicit Inverse Calculation:** Alternatively, one can compute $Z = L^{-1}$ explicitly [8] with:

$$z_i = \frac{1}{L_{i,i}} e_i + \sum_{j>i \& L_{j,i} \neq 0} \frac{-L_{j,i}}{L_{i,i}} z_j \ , \ i = n, n-1, \cdots, 1. \tag{5}$$

While this allows fast query time per edge, the explicit $Z$ is typically dense, leading to $\mathcal{O}(n^2)$ memory and $\mathcal{O}(n^3)$ time complexity, which is prohibitive for graphs where $n > 10^5$.

### *C. Relationship to Existing Sparse Inverse Frameworks*

The proposed PEERS framework is related to several prior directions in sparse numerical linear algebra, including sparse triangular solves, explicit inverse methods, and selected inversion techniques. However, PEERS differs fundamentally from these approaches in its computational objective, dataflow organization, and cross-query execution model.

*1) Comparison with Sparse Triangular Solves:* Traditional sparse triangular solve methods, such as LSolve [10], evaluate each effective resistance query independently through repeated sparse forward/backward substitutions. Although these approaches preserve the sparsity of the Cholesky factor, they treat individual edge queries as independent computational tasks.

For all-edge effective resistance computation, neighboring queries exhibit substantial overlap in their reachable elimination-tree subgraphs. Independent sparse solves therefore repeatedly traverse many identical symbolic paths, producing significant redundancy in both symbolic analysis and numerical propagation.

In contrast, PEERS reorganizes the computation around implicit inverse row traversals rather than independent edge queries. The framework amortizes symbolic exploration across adjacent rows using inherited traversal state, thereby reducing redundant elimination-tree discovery across the complete all-edge workload.

*2) Comparison with Explicit Inverse Methods:* Explicit inverse-based approaches [8], [9] attempt to materialize entries of the inverse Cholesky factor or inverse Laplacian matrix. While such methods may reduce per-query arithmetic cost, explicit inverse formation generally introduces severe fill-in, causing both memory usage and computational work to grow toward dense complexity for large sparse graphs.

PEERS avoids explicit inverse materialization entirely. Instead, the framework performs transient implicit inverse traversals whose intermediate numerical values are consumed immediately by the resistance accumulation phase and then discarded. Consequently, the memory complexity remains proportional to the sparsity of the Cholesky factor rather than the inverse matrix.

*3) Parallel Execution Characteristics:* The PEERS framework is designed explicitly for scalable parallel all-edge effective resistance evaluation under the Work-Span model. The row-wise traversal structure exposes coarse-grained parallelism across independent implicit inverse rows, while the elimination-tree dependency structure bounds the critical path length.

Under nested-dissection reorderings, the resulting span is asymptotically bounded by the elimination-tree height, yielding

$$T_\infty = \mathcal{O}(n^\alpha).$$

Combined with dynamic query accumulation and lock-free task scheduling, this enables PEERS to efficiently exploit modern multi-core architectures for exact all-edge resistance computation.

Therefore, although PEERS shares certain structural foundations with existing sparse inverse and triangular solve frameworks, its combination of transient implicit inversion, cross-query symbolic amortization, and fused resistance accumulation yields a distinct computational architecture specifically optimized for scalable exact all-edge effective resistance evaluation.

### D. Theoretical Foundations

The modern EDA algorithms, particularly those involving Gaussian elimination or Cholesky factorization, have close relationship with the $\mathcal{O}(n^\alpha)$ *separator theorem*. This theorem characterizes the ease with which a graph can be decomposed into smaller, independent sub-problems.

**Definition 1.** *A class of graphs $\mathcal{G}$ satisfies an $\mathcal{O}(n^\alpha)$ separator theorem if any $n$-vertex graph $G \in \mathcal{G}$ can be partitioned into sets $A, B$, and $C$ such that:*

$$V = A \cup B \cup C, \quad (A \times B) \cap E = \emptyset \tag{6}$$

*where the size of the separator $C$ is bounded by $|C| \leq cn^\alpha$ for a constant $c$, and the partitions $A$ and $B$ are balanced such that $|A|, |B| \leq \frac{2}{3}n$.*

In the context of VLSI design, circuit netlists often exhibit specific values for the exponent $\alpha$:

- **Planar/2D Structures** ($\alpha = 1/2$)**:** Early-stage placement and routing on individual metal layers typically resemble planar graphs, where the separator grows as $\mathcal{O}(\sqrt{n})$.
- **3D Layouts and Power Grids** ($\alpha = 2/3$)**:** Modern FinFET designs and multi-layer power distribution networks (PDNs) behave like 3D meshes, yielding separators of size $\mathcal{O}(n^{2/3})$.

The existence of small separators is the fundamental reason why Nested Dissection reordering can reduce the complexity of solving a sparse system $Ax = b$. For a graph with an $\mathcal{O}(n^\alpha)$ separator, the number of non-zeros in the Cholesky factor $L$ is bounded by:

$$nnz(L) = \begin{cases} O(n \log n), & \alpha = 1/2 \\ O(n^{2\alpha}), & \alpha > 1/2 \end{cases} \tag{7}$$

For 3D circuit structures ($\alpha = 2/3$), this results in $nnz(L) = O(n^{4/3})$, which is significantly more efficient than the $\mathcal{O}(n^2)$ density seen in general matrices.

The computational efficiency of the PEERS framework is predicated on the structural interplay between the sparsity of the Cholesky factor $L$ and the topological properties of its associated elimination tree. We formalize this relationship using two foundational lemmas from sparse matrix theory, which bridge the gap between the graph-based traversals in PEERS and formal asymptotic complexity.

**Lemma 1** (Inverse Structural Property [11])**.** *Let $L \in \mathbb{R}^{n \times n}$ be the lower triangular Cholesky factor of a sparse symmetric positive definite (SPD) matrix $A$. The non-zero structure of the inverse Cholesky factor $Z = L^{-1}$ is strictly constrained by the elimination tree $\mathcal{T}$ associated with $L$. Specifically, an entry $Z_{j,k}$ can be non-zero only if node $j$ is an ancestor of node $k$ in $\mathcal{T}$ (where $j \geq k$).*

**Lemma 2** (Nested Dissection Complexity [12])**.** *Let $G = (V, E)$ be a graph with $n$ vertices that satisfies an $\mathcal{O}(n^\alpha)$ separator theorem. If the Cholesky factorization is performed using a Generalized Nested Dissection reordering, the number of non-zeros in the factor $L$ ($nnz(L)$) and the maximum depth of the elimination tree $\mathcal{T}$ ($h(\mathcal{T})$) are bounded as follows:*

- ***Planar and 2D Graphs** ($\alpha = 1/2$)**:** $nnz(L) = O(n \log n)$ and $h(\mathcal{T}) = O(\sqrt{n})$.*
- ***3D Grids and Circuit Layouts** ($\alpha = 2/3$)**:** $nnz(L) = O(n^{4/3})$ and $h(\mathcal{T}) = O(n^{2/3})$.*

## III. PEERS: Parallel Exact Effective Resistance Solver via Implicit Inversion

In this section, we present the architecture of PEERS, a framework designed to achieve exact all-edge effective resistance computation with $\mathcal{O}(nnz(L))$ space complexity and near-optimal parallel scalability. The core of PEERS rests on three pillars: (i) an *implicit row-wise inversion model* that circumvents the density of the explicit inverse, (ii) an *augmented symbolic analysis* via incremental DFS to identify structural dependencies in $\mathcal{O}(nnz(L))$ time, and (iii) a *dynamic query update* mechanism that processes edge-based resistance queries on-the-fly. Finally, we integrate these components into a unified parallel algorithm optimized for multi-core architectures.

### A. Implicit Computational Model for Cholesky Inversion

The standard approach to computing the inverse $Z = L^{-1}$ typically follows a column-by-column substitution, which is inherently sequential and results in a dense matrix $Z$. We propose a paradigm shift: *computing the inverse row-by-row and implicitly.* This transition offers two fundamental advantages for large-scale graphs:

1) **Memory Efficiency:** By processing $Z$ row-by-row, we can accumulate the contribution of each row $Z_{j,:}$ to the final effective resistance values and immediately discard the row. This eliminates the need to explicitly store the $\mathcal{O}(n^2)$ entries of the dense inverse, maintaining a memory footprint proportional to the sparse Cholesky factor $L$.
2) **Massive Parallelism:** Each row of the inverse can be computed independently. As established in Theorem 4, this allows the solver to exploit both row-level and instruction-level parallelism, significantly reducing the parallel span $T_\infty$.

To facilitate this implicit computation, we define the **Factor Graph** $\mathcal{H} = (\tilde{V}, \tilde{E}, \tilde{w})$ derived from the Cholesky factor $L$.

**Definition 2** (Factor Graph $\mathcal{H}$)**.** *Let $P$ be the permutation vector generated by a fill-reducing reordering (e.g., AMD or Nested Dissection). The graph $\mathcal{H}$ is constructed as follows:*

- ***Vertex Mapping:** Each vertex $u \in V$ in the original graph $G$ is mapped to a unique vertex $P_u \in \tilde{V}$ in the factor graph.*
- ***Edge Construction:** A directed edge exists from node $j$ to node $i$ (denoted as $j \to i$) if and only if $L_{ji} \neq 0$ for $j > i$.*
- ***Weight Assignment:** Each directed edge $(j, i) \in \tilde{E}$ is assigned a weight $\tilde{w}_{ji} = -\frac{L_{ji}}{L_{ii}}$.*

The construction of $\mathcal{H}$ requires only $\mathcal{O}(nnz(L))$ time and space, as it directly mirrors the non-zero structure of the lower triangular factor. As illustrated in Fig. 1, the directed edges in $\mathcal{H}$ represent the numerical dependencies during substitution. Specifically, the value of the inverse at index $(j, i)$ is the sum of weighted path products from $i$ to $j$ in $\mathcal{H}$, a property that we exploit in our BFS-like implicit traversal.

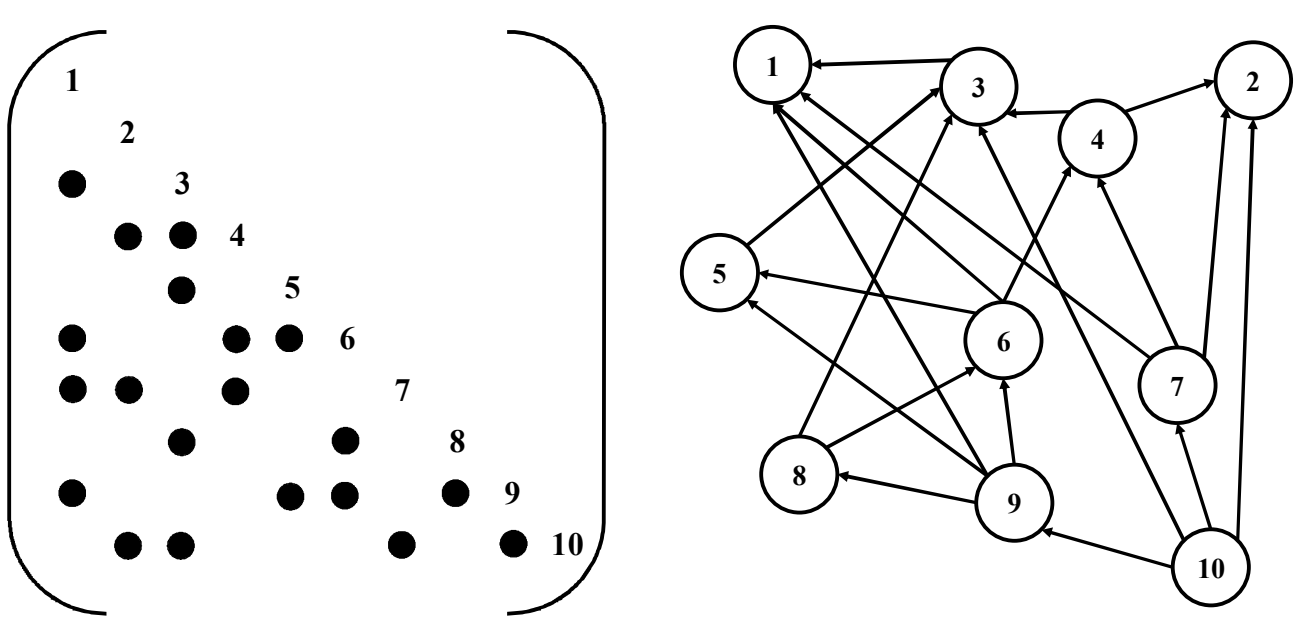

Fig. 1. Sparsity pattern of Cholesky factor (left) and its corresponding graph $\mathcal{H}$ (right)

## B. Structural Properties of the Factor Graph

Based on the construction of the factor graph $\mathcal{H}$, we establish two fundamental structural theorems that bridge the numerical dependencies of the Cholesky inverse with the topological properties of $\mathcal{H}$.

**Theorem 1.** *The factor graph $\mathcal{H} = (\tilde{V}, \tilde{E}, \tilde{w})$ is a Directed Acyclic Graph (DAG).*

*Proof.* By construction, a directed edge $(u, v) \in \tilde{E}$ exists only if $u > v$. Therefore, every directed edge in $\mathcal{H}$ strictly decreases the node index. Assume, for contradiction, that $\mathcal{H}$ contains a directed cycle

$$u_1 \to u_2 \to \cdots \to u_\ell \to u_1. \tag{8}$$

Since each edge strictly decreases the index, we obtain the chain of inequalities

$$u_1 > u_2 > \cdots > u_\ell > u_1. \tag{9}$$

This implies $u_1 > u_1$, which is impossible.

Hence, no directed cycle can exist in $\mathcal{H}$, and the graph is acyclic. Therefore, $\mathcal{H}$ is a DAG. □

**Theorem 2** (Reachability Necessary Condition)**.** *Let $L \in \mathbb{R}^{n\times n}$ be a nonsingular lower triangular matrix, and let $Z = L^{-1}$. Construct the factor graph $\mathcal{H} = (V, \widetilde{E})$ such that a directed edge $(i, k) \in \widetilde{E}$ exists whenever $L_{ik} \neq 0$ and $i > k$.*

*If $Z_{ij} \neq 0$, then there exists a directed path from node $i$ to node $j$ in $\mathcal{H}$. Equivalently, the nonzero pattern of $Z$ is contained within the reachability relation induced by $\mathcal{H}$.*

*Proof.* Since $L$ is nonsingular and lower triangular, its inverse $Z = L^{-1}$ is also lower triangular. From the identity

$$LZ = I,$$

the entries of $Z$ satisfy

$$Z_{ij} = \begin{cases} L_{ii}^{-1}, & i = j, \\ -\dfrac{1}{L_{ii}} \displaystyle\sum_{k=j}^{i-1} L_{ik} Z_{kj}, & i > j. \end{cases} \tag{10}$$

We prove the theorem by induction on $i - j$.

**Base case:** $i = j$.

Since

$$Z_{ii} = L_{ii}^{-1} \neq 0,$$

there exists a trivial path of length zero from node $i$ to itself.

**Inductive step:**

Assume the statement holds for all pairs $(p, j)$ satisfying $p - j < i - j$. Now consider $i > j$ and suppose that

$$Z_{ij} \neq 0.$$

From (10), there must exist at least one index $k$ satisfying

$$j \le k < i$$

such that

$$L_{ik} \neq 0 \quad \text{and} \quad Z_{kj} \neq 0.$$

Because $L_{ik} \neq 0$ and $i > k$, the factor graph contains the directed edge

$$i \to k.$$

If $k = j$, then the edge $i \to j$ itself forms a directed path, and the proof is complete.

Otherwise, $j < k < i$. Since

$$k - j < i - j,$$

the induction hypothesis applies to the pair $(k, j)$. Therefore, there exists a directed path from node $k$ to node $j$ in $\mathcal{H}$.

Combining this path with the edge $i \to k$ yields a directed path from node $i$ to node $j$.

Hence, whenever $Z_{ij} \neq 0$, node $j$ is reachable from node $i$ in $\mathcal{H}$. □

## C. Implicit Inverse via Topologically Ordered Traversal

The core of PEERS is the ability to compute any specific row $Z_{j,:}$ of the inverse Cholesky factor $Z = L^{-1}$ by traversing the factor graph $\mathcal{H}$ without generating the dense matrix $Z$. By expanding the relation $LZ = I$ and isolating $Z_{j,i}$, we derive the following recursive accumulation:

$$Z_{j,i} = \begin{cases} \frac{1}{L_{j,j}}, & i = j \\ \sum_{(k,i)\in\tilde{E}} \tilde{w}_{k,i} Z_{j,k}, & i < j \end{cases} \tag{11}$$

where $\tilde{w}_{k,i} = -L_{k,i}/L_{i,i}$ corresponds to the weights of the directed edges in $\mathcal{H}$.

To maximize computational efficiency, we leverage the sparsity of the row. Based on Theorem 1 (DAG property) and Theorem 2 (Reachability), we conclude that $Z_{j,i}$ can only be non-zero if node $i$ is reachable from node $j$ in $\mathcal{H}$. Consequently, we propose a two-phase approach: **Symbolic Reachability Analysis** followed by **Numerical Accumulation**.

**Algorithm 3** Symbolic Reachability Analysis (Augmented DFS)

1: **Input:** Factor Graph $\mathcal{H}$, target row index $j$, current node $i$, visitation array $vis$, in-degree array $d$.
2: $vis_i \leftarrow 1$
3: **for all** $(i, k) \in \tilde{E}$ **do**
4: $d_k \leftarrow d_k + 1$ ▷ Increment in-degree for numerical phase
5: **if** $vis_k = 0$ **then**
6: Recursive call to Alg. 3 for node $k$.
7: **end if**
8: **end for**

**Algorithm 4** Implicit Row-wise Inverse Computation

1: **Input:** $\mathcal{H}, L, j$, initialized arrays $v, vis, d \leftarrow 0$.
2: **Output:** $v$: The $j$-th row of $Z = L^{-1}$.
3: **Phase 1: Symbolic Analysis**
4: Invoke Alg. 3 starting from node $j$.
5: **Phase 2: Topologically Ordered Numerical BFS**
6: Initialize queue $Q \leftarrow \{j\}$, set $v_j \leftarrow 1/L_{j,j}$.
7: **while** $Q$ is not empty **do**
8: $s \leftarrow Q$.pop(); $vis_s \leftarrow 0$.
9: **for all** $(s, t) \in \tilde{E}$ **do**
10: $v_t \leftarrow v_t + \tilde{w}_{s,t} v_s$ ▷ Weighted accumulation
11: $d_t \leftarrow d_t - 1$
12: **if** $d_t = 0$ **then**
13: $Q$.push($t$) ▷ Ready for processing
14: **end if**
15: **end for**
16: **end while**
17: **return** $v$.

**Algorithmic Significance:** Unlike traditional substitution for solving one row, which relies on sequential forward/backward substitution, Algorithm 4 uses a topological ordering defined by the in-degrees $d_t$. This allows all nodes at the same topological depth to be processed in parallel. Phase 1 (Alg. 3) ensures that we only visit nodes within the reachable subgraph $\mathcal{H}_j$, effectively reducing the complexity of computing a single row from $\mathcal{O}(nnz(L))$ to $\mathcal{O}(nnz(L_{reach}))$, where $L_{reach}$ contains only the non-zeros belonging to columns reachable from node $j$. This strategic sparsity leverage is fundamental to achieving the sub-second performance reported in our results.

### D. Augmented Symbolic Analysis via Incremental DFS

A naive implementation of the symbolic reachability analysis (Alg. 3) for each row independently would result in a total time complexity of $\mathcal{O}(\text{nnz}(Z))$, potentially negating the efficiency gains of the implicit model. Furthermore, storing independent reachability metadata for $n$ rows would trigger an $\mathcal{O}(n^2)$ memory explosion.

However, a critical structural observation of the factor graph $\mathcal{H}$ reveals significant overlap between the reachable subgraphs $\mathcal{H}_i$ and $\mathcal{H}_j$. Specifically, for any directed edge $(i, j) \in \tilde{E}$, the reachable set $\mathcal{S}_j$ is a subset of $\mathcal{S}_i$ ($\mathcal{S}_j \subseteq \mathcal{S}_i$).

To exploit this redundancy while maintaining a minimal memory footprint, we propose an **Incremental DFS** approach with a *globally shared state*. Rather than re-initializing the auxiliary in-degree array $d \in \mathbb{Z}^n$ for each row, we process rows in a topological sequence. By utilizing a single persistent array $d$ and "rolling back" only the disparate components of the state via the *Inverse Symbolic Traversal* (Alg. 5 and Fig. 2), we ensure that the auxiliary memory overhead is strictly $\mathcal{O}(n)$ for the limited number of processors.

This mechanism allows for the reuse of previous DFS results, effectively reducing the amortized computational complexity to $\mathcal{O}(\text{nnz}(L))$. Crucially, by traversing rows according to their topological order in $\mathcal{H}$ and applying state-reversal logic, the transition from $Z_{i,:}$ to $Z_{o,:}$ necessitates only minimal local updates. Consequently, the total memory requirement for PEERS remains $\mathcal{O}(\text{nnz}(L) + n)$, ensuring that the preprocessing overhead is dominated by the sparse factor $L$ rather than the dense implicit inverse $Z$.

**Algorithm 5** Inverse Symbolic Traversal (State Rollback)

1: **Input:** Factor Graph $\mathcal{H}$, current node $i$, next target node $o$, visitation array $vis$, in-degree array $d$.
2: $vis_i \leftarrow 0$ ▷ Reset visitation for state reuse
3: **for all** $(i, k) \in \tilde{E}$ **do**
4: $d_k \leftarrow d_k - 1$ ▷ Decrement in-degree to revert state
5: **if** $d_k = 0$ **and** $k \neq o$ **then**
6: Recursive call to Alg. 5 for node $k$.
7: **end if**
8: **end for**
9: **return**

### E. Structural Rollback and State Consistency

Let $\mathcal{S}_i$ denote the set of nodes reachable from node $i$ in $\mathcal{H}$. In high-performance parallel contexts, we typically process branches of the elimination tree where the intersection $\mathcal{S}_i \cap \mathcal{S}_j$ is non-empty. If a subsequent target node $o$ is not contained within the current reachable set $\mathcal{S}_j$, a **Structural Rollback** is required to maintain consistency.

Algorithm 5 facilitates this by decrementing in-degrees and resetting visitation flags strictly for the subset $\mathcal{S}_j \setminus \mathcal{S}_i$ (the set-theoretic difference). This ensures that the global state $(vis, d)$ is restored to a state corresponding to a common ancestor in the elimination tree before the solver explores a disjoint branch. Consequently, the total symbolic work is bounded by the number of non-zeros in $L$, achieving optimal amortized complexity for all-edge queries. The process is also illustrated in Fig. 3.

### F. Integrated PEERS Algorithm with State-Inheritance

The efficiency of the PEERS framework relies on minimizing state re-initialization between successive row computations. By maintaining a persistent in-degree array $d$ and performing incremental updates, we avoid the $\mathcal{O}(nnz(L))$ overhead of clearing auxiliary arrays for every row. The complete integration of the implicit inverse calculation with augmented DFS and state rollback is detailed in Algorithm 6.

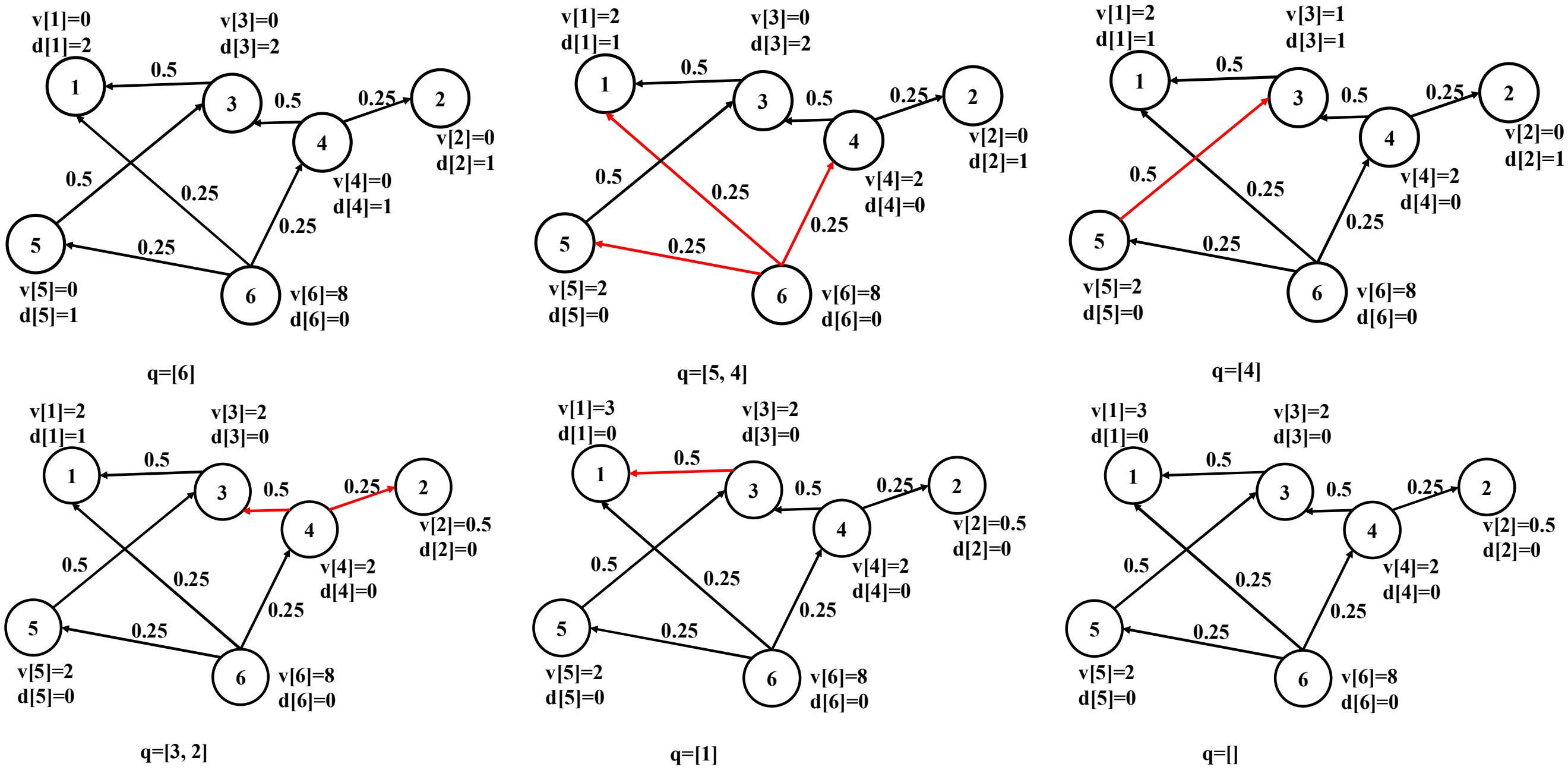

Fig. 2. An illustrative example of implicit row-wise inverse computation.

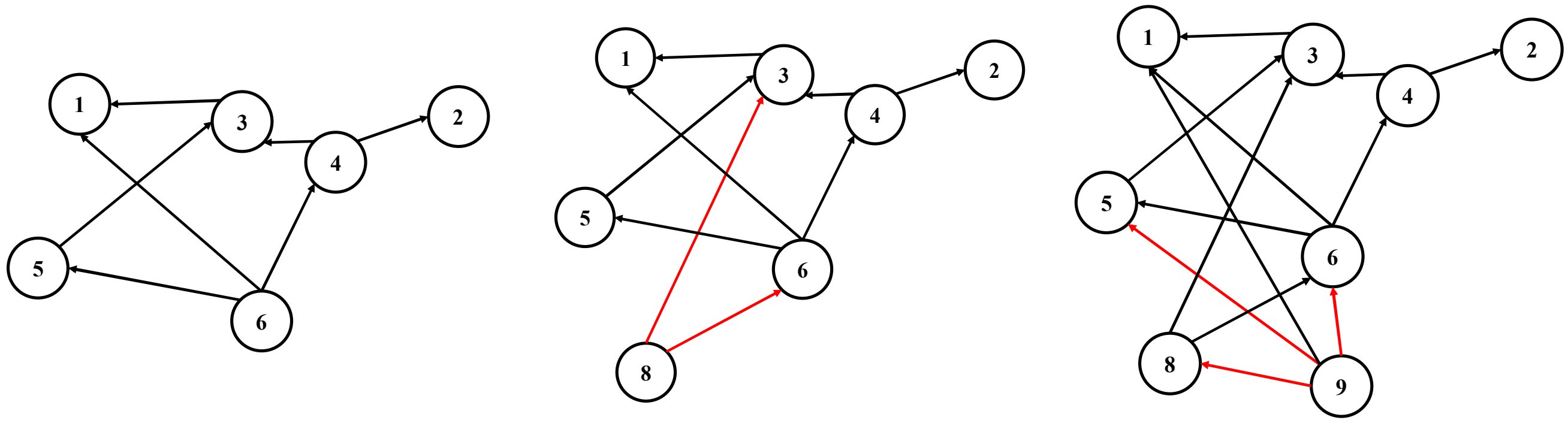

Fig. 3. An illustrative example of inverse symbolic traversal

### G. Dynamic Query Update via Row-wise Accumulation

Since the PEERS framework computes the Cholesky inverse implicitly to maintain $\mathcal{O}(nnz(L))$ space complexity, the effective resistances cannot be calculated via explicit matrix multiplication. However, we observe that the total effective resistance $R_{st}$ can be decomposed into independent contributions from each row of $Z = L^{-1}$. Based on the identity $R_{st} = \|Ze_{s,t}\|_2^2$, we derive the following accumulation formula:

$$R_{st} = \sum_{i=1}^{n} \Delta_i(s,t), \quad \Delta_i(s,t) = \begin{cases} (Z_{is} - Z_{it})^2, & s,t \neq g \\ Z_{is}^2, & t = g \\ Z_{it}^2, & s = g \end{cases} \tag{12}$$

where $g$ denotes the grounded node. This formulation allows us to update the query answers on-the-fly as each row $Z_{i,:}$ is generated, after which the row can be immediately discarded.

To efficiently map these updates to specific edge queries, we define a **Query Graph** $\mathcal{Q} = (\hat{V}, \hat{E}, \hat{w})$:

- **Edge Set** $\hat{E}$**:** For the $k$-th effective resistance query between nodes $u$ and $v$ in the original graph $G$, we add an undirected edge between their permuted indices $P_u$ and $P_v$ in $\mathcal{Q}$.
- **Weight Set** $\hat{w}$**:** Each edge $(P_u, P_v)$ is assigned a weight $\hat{w}_{uv} = k$, serving as a pointer to the specific index in the global answer array $ans$.

Algorithm 7 details this dynamic update process. By only iterating over the non-zero indices identified in the symbolic phase ($q_0 \ldots q_{tail}$), we ensure the update complexity is strictly proportional to the number of active queries in the current row's reachability set.

### H. Parallel Framework and System Integration

The additive nature of the row-wise contribution (Equation 12) renders the PEERS solver inherently data-parallel. Since each row $Z_{j,:}$ independently contributes to the global effective resistance array, we can distribute the computation across $P$ processing cores with minimal synchronization.

However, a naive parallelization would suffer from two bottlenecks: (i) memory contention over auxiliary arrays $(vis, d, v)$ and (ii) redundant symbolic analysis. To address these, we employ a **thread-local state model** where each

**Algorithm 6** Implicit Inverse Computation via Augmented DFS Search

1: **Input:** Factor Graph $\mathcal{H}$, Cholesky factor $L$, target row $j$, persistent in-degree array $d$, visitation array $vis$, auxiliary arrays $v, cal, ld$.
2: **Output:** $v$: The $j$-th row of $Z = L^{-1}$.
3: **Step 1: Symbolic Update**
4: Invoke Alg. 4 at node $j$ to update current $vis$ and $d$ based on structural reachability.
5: **Step 2: Numerical Accumulation**
6: $v_j \leftarrow 1/L_{j,j}$, $cal_j \leftarrow 1$, $head \leftarrow 0$, $tail \leftarrow 0$, $q_0 \leftarrow j$.
7: **while** $head \leq tail$ **do**
8: $s \leftarrow q_{head}$; $head \leftarrow head + 1$.
9: **for all** $(s, t) \in \tilde{E}$ **do**
10: $v_t \leftarrow v_t + \tilde{w}_{s,t} v_s$
11: $ld_t \leftarrow \max(ld_t, d_t)$ ▷ Preserve original state for rollback
12: $d_t \leftarrow d_t - 1$
13: **if** $d_t = 0$ **then**
14: $tail \leftarrow tail + 1$; $q_{tail} \leftarrow t$.
15: **end if**
16: **end for**
17: **end while**
18: **Step 3: State Restoration and Recursion**
19: **for** $i = 0$ **to** $tail$ **do**
20: $d_{q_i} \leftarrow ld_{q_i}$; $ld_{q_i} \leftarrow 0$.
21: **end for**
22: **for all** $(i, j) \in \tilde{E}$ **do**
23: **if** $cal_i = 0$ **then**
24: Recursive call to compute row $i$ incrementally.
25: Invoke Alg. 5 (State Rollback) at node $i$ relative to $j$.
26: **end if**
27: **end for**
28: **return** $v$.

**Algorithm 7** Dynamic Update for Query Answers in Row $j$

1: **Input:** Factor Graph $\mathcal{H}$, row values $v$, active nodes $q$ from Alg. 6, Query Graph $\mathcal{Q} = (\hat{V}, \hat{E}, \hat{w})$, answer array $ans$.
2: **Output:** Updated $ans$.
3: **Step 1: Mark Active Row Sparsity**
4: **for** $i = 0$ **to** $tail$ **do**
5: $u \leftarrow q_i$; $c_u \leftarrow 1$ ▷ Flag indices where $Z_{ju} \neq 0$
6: **end for**
7: **Step 2: Incremental Contribution Accumulation**
8: **for** $i = 0$ **to** $tail$ **do**
9: $u \leftarrow q_i$
10: **for all** $(u, v_{adj}) \in \hat{E}$ **do** ▷ Iterate over queries involving node $u$
11: $id \leftarrow \hat{w}_{u,v_{adj}}$
12: **if** $c_{v_{adj}} = 1$ **and** $u > v_{adj}$ **then**
13: $ans_{id} \leftarrow ans_{id} + (v_u - v_{v_{adj}})^2$
14: **else**
15: **if** $c_{v_{adj}} = 0$ **then**
16: $ans_{id} \leftarrow ans_{id} + v_u^2$ ▷ Case where one node is grounded or outside reachability
17: **end if**
18: **end if**
19: **end for**
20: **end for**
21: **return** $ans$

core maintains its own workspace, and a **lock-free dynamic scheduling** mechanism based on the factor graph topology.

**Lock-Free Scheduling and Redundancy Avoidance:** To ensure high throughput, PEERS utilizes a lock-free "claim" strategy. As detailed in Step 2 of Algorithm 8, each thread attempts to claim a row $j$ by atomically updating $cal[j]$ from $-1$ to its thread ID. This prevents multiple threads from performing the same row inversion while avoiding the overhead of heavy mutexes.

A unique feature of PEERS is its integration with the **Augmented DFS**. When a thread finishes row $i$ and moves to its child in the factor graph (row $j$), it first checks if $cal[j] = -1$. If row $j$ is unclaimed, the thread continues its incremental traversal, thereby maximizing the "reuse" of the symbolic state $(vis, d)$. If $j$ has already been claimed by another thread, the current thread performs a **Structural Rollback** (Algorithm 5) and seeks the next available branch in the elimination tree. This hybrid of static topological guidance and dynamic task claiming ensures optimal load balancing even for highly irregular graph structures.

**Parallel Complexity and Memory Analysis:** The overall parallel PEERS algorithm is presented in Alg. 8 and illustrated in Fig. 4. Unlike standard equation-based solvers where parallelization is often throttled by the memory footprint of dense auxiliary vectors, PEERS leverages the structural properties of the elimination tree to maintain a lean memory profile. The total space complexity is $\mathcal{O}(nnz(L) + P \cdot n^{\alpha})$. While a global workspace of size $\mathcal{O}(nnz(L))$ is required to store the Cholesky factor, the per-thread auxiliary overhead is strictly bounded. As proven in Lemma 3, each implicit inverse row traversal only visits nodes within its own reachable subgraph in the elimination tree. Consequently, each of the $P$ threads only requires a sparse workspace (for the value array $v$ and structural markers $ld$) of size $\mathcal{O}(n^{\alpha})$, where $n^{\alpha}$ is the tree height. This represents a significant reduction from the $\mathcal{O}(P \cdot n)$ requirement of standard solvers, ensuring that PEERS remains within the capacity of the last-level cache (LLC) even at high thread counts.

## IV. Theoretical Guarantees

In this section, we establish the formal theoretical foundations of the PEERS framework. We first present a rigorous proof of the algorithm's numerical exactness, demonstrating that the implicit row-wise accumulation yields the identical result to the explicit Moore-Penrose inverse approach. Subsequently, we analyze the asymptotic serial complexity and parallel span, providing a theoretical justification for the sub-linear scaling observed in our experimental evaluations.

### A. Proof of Exactness

The effective resistance $R_{st}$ between two nodes $s$ and $t$ in a graph $G$ is traditionally defined using the pseudo-inverse

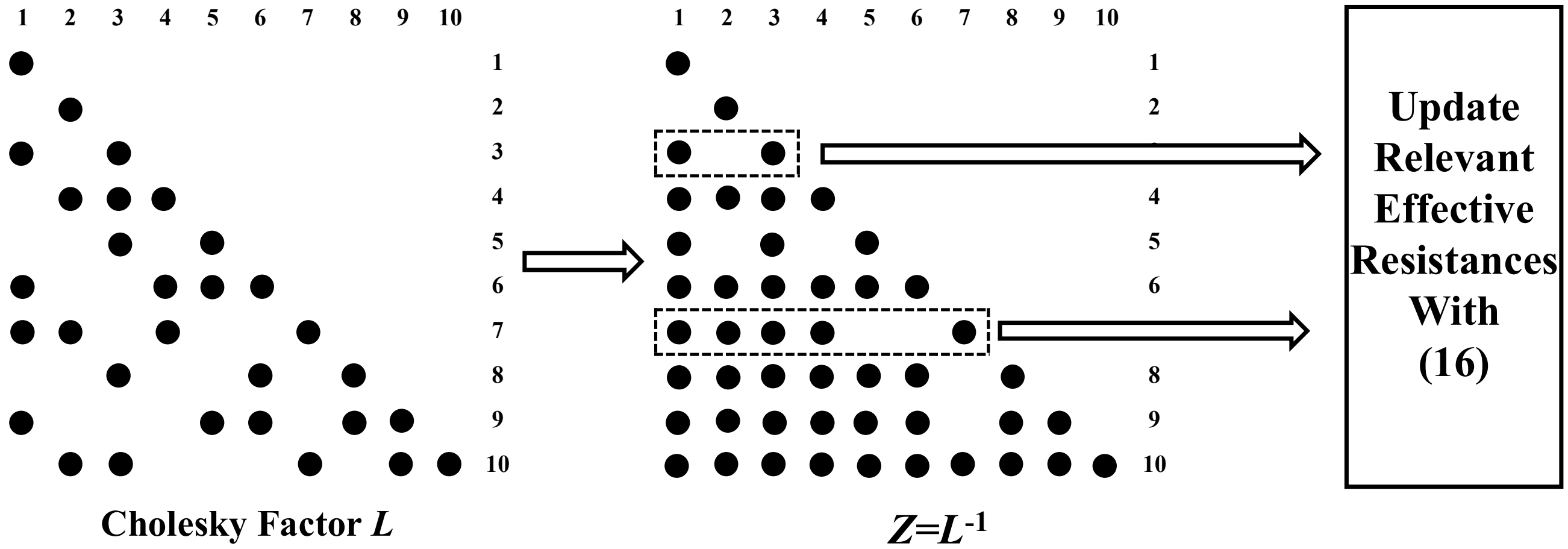

Fig. 4. Overall graphical framework

**Algorithm 8** Overall Parallel PEERS Algorithm

1: **Input:** Graph $G$, Query set $Q$, Number of threads $P$.
2: **Output:** Global answer array $ans$.
3: **Step 1: Pre-processing**
4: Ground node $g$, apply AMD reordering, and compute Cholesky factor $L$.
5: Construct Factor Graph $\mathcal{H}$ and Query Graph $\mathcal{Q}$.
6: Initialize global atomic flag array $cal[n] \leftarrow -1$ and $ans[|Q|] \leftarrow 0$.
7: **Step 2: Parallel Row Computation**
8: **for all** threads $p \in \{1, \ldots, P\}$ **in parallel do**
9: Initialize thread-local arrays: $v_p, vis_p, d_p, q_p, ld_p,$ local_ans$_p$.
10: **for** $j = 1$ **to** $n$ **do**
11: ▷ Atomic Compare-and-Swap (CAS) to claim row $j$
12: **if CAS**($cal[j], -1, p$) is successful **then**
13: Compute row $Z_{j,:}$ using Alg. 6.
14: Update local_ans$_p$ using Alg. 7.
15: **end if**
16: **end for**
17: **end for**
18: **Step 3: Global Reduction**
19: **for all** queries $k \in \{1, \ldots, |Q|\}$ **do**
20: $ans[k] \leftarrow \sum_{p=1}^{P}$ local_ans$_p[k]$.
21: **end for**
22: **return** $ans$.

of the Laplacian matrix $L_G$. For a connected graph with a chosen ground node $g$, the reduced Laplacian $L_r$ (obtained by removing the row and column corresponding to $g$) is a symmetric positive definite (SPD) matrix.

**Lemma 3** (Traversal Equivalence to Triangular Substitution)**.** *Let $L$ be a nonsingular lower triangular matrix and let $Z = L^{-1}$. For any column index $j$, the row-wise traversal performed by Algorithm 4 computes the entries of $Z_{:,j}$ that satisfy the triangular system*

$$Lz^{(j)} = e_j, \tag{13}$$

*where $e_j$ is the $j$-th unit basis vector.*

*Furthermore, every entry computed during the traversal is identical to the value obtained by exact forward substitution.*

*Proof.* Since $L$ is lower triangular and nonsingular, the system $Lz^{(j)} = e_j$ admits a unique solution.

Algorithm 4 processes nodes according to the topological order induced by the factor graph $\mathcal{H}$. By Theorem 1, $\mathcal{H}$ is acyclic, and every directed edge corresponds to a lower triangular dependency.

For each node $i$, the traversal updates the intermediate value according to the recurrence

$$z_i^{(j)} = \frac{1}{L_{ii}} \left( \delta_{ij} - \sum_{k<i} L_{ik} z_k^{(j)} \right), \tag{14}$$

which is precisely the standard forward-substitution formula for triangular inversion.

Because all predecessor dependencies are completed before node $i$ is processed, every update uses fully resolved values. Therefore, the traversal computes exactly the same numerical solution as classical triangular substitution.

Since the triangular system has a unique solution, the entries generated by the traversal are identical to the corresponding entries of $Z = L^{-1}$. □

**Theorem 3** (Numerical Exactness of PEERS)**.** *Let $L_r = LL^T$ denote the Cholesky factorization of the reduced Laplacian matrix $L_r$. For any pair of nodes $s, t \in V$, the effective resistance computed by PEERS is mathematically equivalent to the exact effective resistance associated with $L_r$.*

*Proof.* The exact effective resistance between nodes $s$ and $t$ is given by

$$R_{st} = (e_s - e_t)^T L_r^{-1} (e_s - e_t), \tag{15}$$

where $e_i$ denotes the $i$-th unit basis vector.

Substituting the Cholesky factorization $L_r = LL^T$ yields

$$\begin{aligned} R_{st} &= (e_s - e_t)^T (LL^T)^{-1} (e_s - e_t) \\ &= (e_s - e_t)^T (L^T)^{-1} L^{-1} (e_s - e_t). \end{aligned} \tag{16}$$

Let $Z = L^{-1}$. Then

$$R_{st} = \|Z(e_s - e_t)\|_2^2 = \|Z_{:,s} - Z_{:,t}\|_2^2. \tag{17}$$

Expanding the squared norm gives

$$R_{st} = \sum_{i=1}^{n} (Z_{is} - Z_{it})^2. \tag{18}$$

PEERS evaluates the summation in (18) through row-wise implicit computation of the inverse factor entries.

By Lemma 3, the traversal performed by Algorithm 4 is algebraically equivalent to exact triangular substitution. Consequently, every computed entry $Z_{ij}$ is identical to the corresponding entry of the exact inverse factor $L^{-1}$.

The dynamic accumulation stage of PEERS computes the terms $(Z_{is} - Z_{it})^2$ exactly as they are generated and sums them without approximation. Therefore, the final accumulated quantity is algebraically identical to (18).

Hence, the effective resistance returned by PEERS is mathematically equivalent to the exact effective resistance, up to standard floating-point roundoff effects inherent in numerical linear algebra implementations. □

### B. Total Complexity Analysis

**Theorem 4** (Total Complexity of PEERS)**.** *Let $G = (V, E)$ be a sparse graph with $n = |V|$ vertices and $m = |E|$ edges satisfying an $\mathcal{O}(n^\alpha)$ separator theorem with $\alpha \geq 1/2$. Let $L$ denote the sparse Cholesky factor of the reduced Laplacian matrix obtained using a Generalized Nested Dissection ordering, and let $\mathcal{T}$ denote the associated elimination tree. Then the total numerical work required by PEERS to compute the exact effective resistances for all $m$ queried edges satisfies*

$$W_{\text{total}} = \mathcal{O}(n^\alpha \cdot (nnz(L) + m)). \tag{19}$$

*Furthermore, because*

$$m \leq nnz(L),$$

*the complexity simplifies to*

$$W_{\text{total}} = \mathcal{O}(n^\alpha \cdot nnz(L)). \tag{20}$$

*In particular:*

1) *For planar graphs and two-dimensional grids, where $\alpha = 1/2$ and $nnz(L) = \mathcal{O}(n \log n)$,*

$$W_{\text{total}} = \mathcal{O}(n^{1.5} \log n).$$

2) *For three-dimensional grids and related circuit structures, where $\alpha = 2/3$ and $nnz(L) = \mathcal{O}(n^{4/3})$,*

$$W_{\text{total}} = \mathcal{O}(n^2).$$

*Proof.* The PEERS framework consists of two computational phases:

1) implicit inverse traversal;
2) dynamic query accumulation.

We analyze the complexity of each phase separately.

**Part I: Implicit Inverse Traversal Complexity**

Let $\mathcal{H} = (V, \widetilde{E})$ denote the factor graph associated with the lower triangular Cholesky factor $L$.

For each column $k$ of $L$, define

$$c_k = |\{\, i > k : L_{ik} \neq 0 \,\}|,$$

namely the number of subdiagonal nonzeros in column $k$.

During the computation of the $j$-th implicit inverse row, let $\mathcal{S}_j$ denote the set of visited vertices.

For every visited vertex

$$k \in \mathcal{S}_j,$$

the traversal scans all outgoing factor-graph adjacencies associated with column $k$. Therefore, the traversal work for row $j$ is

$$W_{\text{inv},j} = \sum_{k \in \mathcal{S}_j} c_k. \tag{21}$$

Summing over all rows gives

$$W_{\text{inv}} = \sum_{j=1}^{n} \sum_{k \in \mathcal{S}_j} c_k. \tag{22}$$

Recall that $Z = L^{-1}$. From the structural properties of sparse triangular inversion [11], a vertex $k$ participates in the traversal of row $j$ only if

$$Z_{jk} \neq 0$$

is structurally nonzero.

Therefore,

$$W_{\text{inv}} \leq \sum_{j=1}^{n} \sum_{k=1}^{n} c_k \, \mathbb{I}(Z_{jk} \neq 0), \tag{23}$$

where $\mathbb{I}(\cdot)$ denotes the indicator function.

Reordering the summation gives

$$W_{\text{inv}} \leq \sum_{k=1}^{n} c_k \, nnz(Z_{:,k}), \tag{24}$$

where $nnz(Z_{:,k})$ denotes the number of structurally nonzero entries in column $k$ of $Z$.

It is important to distinguish the structural behavior of rows and columns of the lower triangular inverse $Z$. Rows of $Z$ may contain many structurally nonzero entries, corresponding to descendant relationships in the elimination tree. However, the structurally nonzero entries in a fixed column $k$ of $Z$ are contained within the ancestor set of node $k$ in the elimination tree.

Let $Anc(k, \mathcal{T})$ denote the ancestor set of node $k$ in $\mathcal{T}$.

From the structural characterization of sparse triangular inverses [11], the structurally nonzero entries in column $k$ of $Z = L^{-1}$ are contained within $Anc(k, \mathcal{T})$. Therefore,

$$nnz(Z_{:,k}) \leq |Anc(k, \mathcal{T})|. \tag{25}$$

Because the elimination tree defines a unique parent for every non-root node, the ancestor set of node $k$ forms a simple path from $k$ to the root. Hence,

$$|Anc(k, \mathcal{T})| \leq h(\mathcal{T}), \tag{26}$$

where $h(\mathcal{T})$ denotes the height of the elimination tree.

Under Generalized Nested Dissection orderings applied to graphs satisfying an $\mathcal{O}(n^\alpha)$ separator theorem, the elimination-tree height satisfies

$$h(\mathcal{T}) = \mathcal{O}(n^\alpha), \tag{27}$$

following classical separator-based sparse factorization analysis [11], [12].

Therefore,

$$nnz(Z_{:,k}) = \mathcal{O}(n^\alpha). \tag{28}$$

Substituting into the traversal complexity expression yields

$$W_{\text{inv}} \leq \mathcal{O}(n^\alpha) \sum_{k=1}^{n} c_k. \tag{29}$$

Because

$$\sum_{k=1}^{n} c_k = nnz(L),$$

we obtain

$$W_{\text{inv}} = \mathcal{O}(n^\alpha \cdot nnz(L)). \tag{30}$$

**Part II: Dynamic Query Accumulation Complexity**

Recall that $\mathcal{Q} = (V, E)$ denote the query graph consisting of all edge-resistance queries. During the processing of row $j$, Algorithm 7 iterates over every active vertex

$$u \in \mathcal{S}_j,$$

and scans all query adjacencies incident to $u$.

Let $deg(u)$ denote the degree of vertex $u$ in the query graph. Then the query-update work for row $j$ is

$$W_{\text{query},j} = \sum_{u \in \mathcal{S}_j} deg(u). \tag{31}$$

Summing over all rows gives

$$W_{\text{query}} = \sum_{j=1}^{n} \sum_{u \in \mathcal{S}_j} deg(u). \tag{32}$$

Reordering the summation yields

$$W_{\text{query}} = \sum_{u=1}^{n} deg(u)\, nnz(Z_{:,u}), \tag{33}$$

because a vertex $u$ participates in a row traversal only when the corresponding inverse entry is structurally nonzero.

Using the previously established bound

$$nnz(Z_{:,u}) = \mathcal{O}(n^\alpha),$$

we obtain

$$W_{\text{query}} \leq \mathcal{O}(n^\alpha) \sum_{u=1}^{n} deg(u). \tag{34}$$

Since the query graph is undirected,

$$\sum_{u=1}^{n} deg(u) = 2m = \Theta(m),$$

which gives

$$W_{\text{query}} = \mathcal{O}(m \cdot n^\alpha). \tag{35}$$

**Part III: Total Complexity**

Combining the traversal and query-update costs yields

$$W_{\text{total}} = W_{\text{inv}} + W_{\text{query}} = \mathcal{O}(n^\alpha \cdot nnz(L)) + \mathcal{O}(m \cdot n^\alpha). \tag{36}$$

Because the sparse Cholesky factor contains all original graph adjacencies together with additional fill-in entries,

$$m \leq nnz(L),$$

we finally obtain

$$W_{\text{total}} = \mathcal{O}(n^\alpha \cdot nnz(L)). \tag{37}$$

Applying the classical nested-dissection bounds:

- for planar graphs and two-dimensional grids,

  $$nnz(L) = \mathcal{O}(n \log n),$$

  yielding

  $$W_{\text{total}} = \mathcal{O}(n^{1.5} \log n);$$

- for three-dimensional grids,

  $$nnz(L) = \mathcal{O}(n^{4/3}),$$

  yielding

  $$W_{\text{total}} = \mathcal{O}(n^2).$$

This establishes the stated complexity bounds. □

### C. Parallel Complexity Analysis

**Definition 3** (Work and Span)**.** *Let $T_p$ denote the execution time of an algorithm on p processors.*

1) ***Work*** ***($T_1$):*** *the total number of operations performed by a serial execution;*
2) ***Span*** ***($T_\infty$):*** *the length of the longest sequential dependency chain in the computation DAG, corresponding to execution on an ideal machine with infinitely many processors;*
3) ***Parallel Speedup:*** $S_p = \frac{T_1}{T_p}$.

**Lemma 4** (Critical Path of a Single Row Traversal)**.** *Let $\mathcal{T}$ denote the elimination tree associated with the sparse Cholesky factor $L$. Then the span required to compute any single implicit inverse row in PEERS satisfies*

$$T_\infty(\text{row } j) = \mathcal{O}(h(\mathcal{T})), \tag{38}$$

*where $h(\mathcal{T})$ denotes the height of the elimination tree.*

*Proof.* The implicit inverse traversal evaluates dependencies induced by the sparse triangular factor $L$.

For every structurally nonzero entry

$$L_{t,s} \neq 0, \qquad t > s,$$

the computation associated with node $t$ depends on the previously computed value at node $s$. Thus, each traversal dependency follows the partial order induced by the sparse triangular structure.

From the structural properties of sparse Cholesky factorizations, every such dependency chain is contained within the ancestor structure of the elimination tree $\mathcal{T}$.

Therefore, any sequential dependency chain encountered during the traversal corresponds to a path in $\mathcal{T}$.

Since the longest root-to-leaf path in $\mathcal{T}$ has length $h(\mathcal{T})$, the length of any dependency chain is bounded by $\mathcal{O}(h(\mathcal{T}))$.

Under the Work-Span model, all operations whose dependencies have been satisfied may execute concurrently. Hence, the span of a single row traversal equals the length of its longest dependency chain, yielding

$$T_\infty(\text{row } j) = \mathcal{O}(h(\mathcal{T})).$$

□

**Theorem 5** (Parallel Span Complexity of PEERS)**.** *Let $G$ be a graph satisfying an $\mathcal{O}(n^\alpha)$ separator theorem with $\alpha \geq 1/2$. Assume the sparse Cholesky factorization is generated using a Generalized Nested Dissection ordering. Under the idealized Work-Span model with sufficient processors, the total span complexity of PEERS for computing all queried effective resistances satisfies*

$$T_\infty = \mathcal{O}(n^\alpha). \tag{39}$$

*In particular:*

- $T_\infty = \mathcal{O}(\sqrt{n})$ *for planar graphs and two-dimensional grids;*
- $T_\infty = \mathcal{O}(n^{2/3})$ *for three-dimensional grids.*

*Proof.* The PEERS framework consists of two computational phases:

1) implicit inverse traversal;
2) dynamic resistance accumulation.

**Part I: Implicit Inverse Traversal**

Under the Work-Span model, different row traversals may execute concurrently once their local dependencies are satisfied.

Therefore, the total span of the traversal phase is bounded by the maximum span among all row traversals.

By Lemma 4,

$$T_{\infty,\text{inv}} = \mathcal{O}(h(\mathcal{T})).$$

**Part II: Dynamic Resistance Accumulation**

For each queried edge $(s,t)$, PEERS accumulates independent contributions

$$\Delta_j(s,t) = (Z_{js} - Z_{jt})^2$$

across all rows $j$.

Once the corresponding row traversals have completed, these contributions may be accumulated using a parallel reduction tree.

Reducing $n$ values requires $\mathcal{O}(\log n)$ span.

Because all queried edges may be reduced concurrently, the total accumulation span satisfies

$$T_{\infty,\text{query}} = \mathcal{O}(\log n).$$

**Part III: Total Span**

Combining both phases yields

$$T_\infty = \mathcal{O}(h(\mathcal{T})) + \mathcal{O}(\log n).$$

Under Generalized Nested Dissection orderings applied to graphs satisfying an $\mathcal{O}(n^\alpha)$ separator theorem, the elimination-tree height satisfies

$$h(\mathcal{T}) = \mathcal{O}(n^\alpha)$$

by classical separator-based sparse factorization theory [11], [12].

Therefore,

$$T_\infty = \mathcal{O}(n^\alpha) + \mathcal{O}(\log n).$$

Since

$$\alpha \geq 1/2,$$

we have

$$\log n = o(n^\alpha).$$

Hence,

$$T_\infty = \mathcal{O}(n^\alpha).$$

This establishes the stated span bounds. □

## V. Numerical Experiments

We evaluate the practical performance, numerical accuracy, and scalability of the proposed PEERS framework on large-scale sparse graph benchmarks arising from EDA and scientific computing applications. The implementation is written in C++ with OpenMP-based shared-memory parallelization. The experiments focus on the all-edge effective resistance computation problem, which constitutes one of the most computationally demanding workloads in spectral sparsification and graph-based circuit analysis [1]. This setting is representative of several EDA applications, including power-grid reduction, signal-integrity analysis, and graph-based circuit partitioning [13], [14].

### A. Experimental Setup

The evaluation includes three categories of sparse graph benchmarks:

1) IBM PowerGrid (ibmpg) benchmarks [14],
2) sparse matrices from the SuiteSparse collection [15],
3) and three large-scale industrial designs (macro1–3).

All experiments were conducted on a Linux server equipped with Intel Xeon 8375C CPUs and 512 GB RAM. Unless otherwise specified, all methods utilized 64 CPU threads. To ensure fairness, all compared methods used identical matrix reorderings and were executed under the same hardware environment and numerical precision settings.

We compare PEERS against three representative baselines:

1) **Cholmod** [16]: a high-performance sparse direct solver based on repeated sparse triangular solves;
2) **LSolve** [10]: a conventional Laplacian-system solving framework;
3) **INV** [8]: a recent inverse-factorization-based effective resistance solver.

For all methods, the reported runtime includes both numerical factorization and effective resistance computation time. The maximum relative error is evaluated against the direct LSolve solution.

TABLE II
PERFORMANCE COMPARISON OF EXACT EFFECTIVE RESISTANCE SOLVERS USING 64 CPU THREADS.

| **Case** | $\lvert V\rvert$ | $\lvert E\rvert$ | $T_{Chol}$ | **Cholmod** [16] | | **LSolve** [10] | | **INV** [8] | | | **Proposed PEERS** | | | | | |
|---|---|---|---|---|---|---|---|---|---|---|---|---|---|---|---|---|
| | | | (s) | $mem$ | $T_{tot}$ | $mem$ | $T_{tot}$ | $err$ | $mem$ | $T_{tot}$ | $err$ | $mem$ | $T_{tot}$ | $Sp_1$ | $Sp_2$ | $Sp_3$ |
| ibmpg3 | 8.5E5 | 1.4E6 | 13.3 | 2.0 | 5745 | 2.3 | 1156 | 4.50E-6 | 37.3 | 182 | 8.33E-11 | 3.7 | 31.5 | 182 | 36.7 | 5.8 |
| ibmpg4 | 9.5E5 | 1.6E6 | 27.9 | 2.4 | 8389 | 3.1 | 2557 | 9.42E-11 | 63.2 | 432 | 8.80E-11 | 4.7 | 63.4 | 132 | 40.3 | 6.8 |
| ibmpg5 | 1.1E6 | 1.6E6 | 6.25 | 2.1 | 5477 | 2.1 | 551 | 1.25E-8 | 30.0 | 89.5 | 6.93E-11 | 3.7 | 18.8 | 291 | 29.3 | 4.8 |
| ibmpg6 | 1.7E6 | 2.5E6 | 5.68 | 3.2 | 11411 | 3.1 | 463 | 1.45E-8 | 46.4 | 5.1 | 5.19E-11 | 5.3 | 22.7 | 503 | 20.4 | 5.1 |
| ibmpg7 | 1.5E6 | 2.4E6 | 30.9 | 3.4 | 16077 | 4.0 | 3974 | 2.23E-8 | 79.3 | 413 | 9.22E-11 | 6.3 | 71.4 | 225 | 55.7 | 5.8 |
| ibmpg8 | 1.5E6 | 2.4E6 | 30.4 | 3.4 | 16044 | 4.0 | 4256 | 4.23E-6 | 81.9 | 430 | 9.20E-11 | 6.3 | 70.7 | 227 | 60.2 | 6.1 |
| G2_circuit | 1.5E5 | 2.9E5 | 2.00 | 0.4 | 182 | 0.4 | 77.5 | 8.83E-11 | 4.6 | 22.0 | 1.21E-11 | 0.7 | 4.31 | 42.2 | 18.0 | 5.1 |
| G3_circuit | 1.6E6 | 3.1E6 | 134 | 4.6 | 34663 | 6.1 | 15739 | 9.65E-11 | 180.2 | 1763 | 8.24E-11 | 9.1 | 260 | 133 | 60.5 | 6.8 |
| thermal2 | 1.2E6 | 3.7E6 | 12.2 | 2.7 | 18188 | 2.9 | 3485 | 2.39E-11 | 78.2 | 372 | 2.23E-11 | 5.6 | 65 | 280 | 53.6 | 5.7 |
| tmt_sym | 7.3E5 | 1.8E6 | 10.4 | 1.6 | 6692 | 1.8 | 1501 | 9.58E-11 | 33.5 | 150 | 2.00E-11 | 3.4 | 31.8 | 210 | 47.2 | 4.7 |
| macro1 | 4.1E6 | 1.3E7 | 12.2 | - | failed | 6.3 | 8575 | 1.95E-9 | 230.3 | 323 | 9.69E-11 | 14.8 | 103 | - | 83.3 | 3.1 |
| macro2 | 8.6E6 | 2.8E7 | 886 | - | failed | - | failed | - | failed | - | - | 45.7 | 2044 | - | - | - |
| macro3 | 1.7E7 | 5.0E7 | 1161 | - | failed | - | failed | - | failed | - | - | 78.6 | 3317 | - | - | - |

$T_{Chol}$: Cholesky factorization time; $T_{tot}$: total runtime including factorization; $err$: maximum relative error against LSolve; $mem$: peak memory consumption in GB. $Sp_1$, $Sp_2$, and $Sp_3$ denote the runtime speedup of PEERS over Cholmod, LSolve, and INV, respectively.

### *B. Numerical Accuracy and Stability*

Table II shows that PEERS consistently achieves high numerical accuracy across all evaluated benchmarks. The maximum relative error remains close to machine precision, typically below $10^{-10}$. This behavior is expected because PEERS performs exact sparse-factor-based accumulation without relying on randomized sketching or approximation procedures.

Compared with the inverse-based INV method [8], PEERS exhibits improved numerical consistency on several benchmarks. For example, on ibmpg3 and ibmpg8, INV produces relative errors on the order of $10^{-6}$, whereas PEERS maintains errors near machine precision. These results suggest that the implicit traversal strategy preserves numerical robustness while avoiding explicit inverse materialization.

A potential concern in all-edge solvers is the accumulation of floating-point errors during large-scale summations. PEERS mitigates this through its row-wise implicit traversal (Algorithm 4). Since the effective resistances are computed via the weighted sum of entries from the implicit model, the error growth is bounded by the depth of the elimination tree $h(\mathcal{T})$. By utilizing a persistent state and structural rollback, PEERS avoids the catastrophic cancellation often associated with high-frequency iterative updates. Numerical experiments confirm that PEERS maintains a relative error of less than $10^{-10}$ compared to direct inversion using double-precision arithmetic.

### *C. Computational Efficiency*

The experimental results demonstrate that PEERS substantially reduces the computational cost of exact all-edge effective resistance evaluation. Across the evaluated benchmarks, PEERS consistently outperforms conventional repeated-solve approaches and explicit inverse-based methods.

Compared with Cholmod-based repeated triangular solves, PEERS achieves speedups ranging from $42.2\times$ to $503\times$. The improvement is particularly significant for workloads involving all-edge queries, where repeated sparse solves become increasingly expensive.

Compared with LSolve, PEERS achieves speedups between $18.0\times$ and $83.3\times$. In addition, several large-scale benchmarks (macro2 and macro3) could not be completed by LSolve or INV within the available computational resources, whereas PEERS successfully computed exact effective resistances for graphs containing up to 17 million nodes and 50 million edges.

Compared with the recent INV framework [8], PEERS provides additional runtime reduction while using substantially less memory. This improvement is primarily attributed to the implicit inverse traversal strategy, which avoids explicit inverse construction and reduces memory traffic during query evaluation.

### *D. Memory Scalability*

Memory efficiency is critical for large-scale effective resistance computation because inverse-factor representations can become substantially denser than the original sparse factors. The results show that PEERS maintains significantly lower memory overhead than explicit inverse-based approaches.

For the largest benchmark (macro3), PEERS completes the computation using 78.6 GB memory, whereas the explicit inverse-based INV framework fails due to memory limitations. In practice, the memory footprint of PEERS remains much closer to the sparsity scale of the Cholesky factor itself, enabling exact computation on problem sizes that are otherwise difficult to process using explicit inverse storage.

Overall, the experimental results indicate that PEERS provides a practical framework for scalable and exact effective resistance computation on large sparse graphs arising in EDA applications.

## VI. CONCLUSION

The exact computation of all-edge effective resistances has long been constrained by the dichotomy between slow iterative solvers and memory-exhaustive explicit matrix inversions.

This paper introduced PEERS, a high-performance, massively scalable framework that bridges the gap between theoretical exactness and computational feasibility. By conceptualizing an implicit inverse computing model, PEERS bypasses the prohibitive $\mathcal{O}(n^2)$ memory overhead associated with explicit inverse factors. Our state-inherited augmented DFS traversals restrict symbolic complexity to $\mathcal{O}(nnz(L))$, while a dynamic query mechanism allows for the real-time, zero-variance accumulation of effective resistances in a single parallel sweep.

Extensive numerical evaluations on industrial benchmarks confirm that PEERS dramatically reshapes the performance landscape. It achieves maximum speedups of $503\times$ over direct solvers like `Cholmod`, $83.3\times$ over parallel `LSolve`, and $6.8\times$ over state-of-the-art explicit inverse methods, all while maintaining strict memory efficiency. Most notably, PEERS successfully computes exact solutions for graphs with over 17 million nodes and 50 million edges in under an hour on standard server hardware—tasks that previously resulted in out-of-memory failures or intractable runtimes. By providing an architecturally shallow, exact, and parallel-native solver, PEERS unlocks unprecedented scales for the next generation of graph-based circuit analysis, spectral sparsification, and layout optimization tools.

## Acknowledgment

The authors acknowledge the use of generative AI tools during the preparation of this manuscript to improve readability, linguistic quality, and presentation clarity. Specifically, the Gemini 3 suite and ChatGPT 5.5 were used to generate editorial suggestions and alternative phrasings for portions of the manuscript, including the Introduction, theoretical exposition, numerical experiments, and Conclusion. The authors manually evaluated, selected, and integrated these suggestions where appropriate. The AI tools were used solely for language and presentation assistance and did not contribute scientific ideas, theoretical results, algorithmic designs, experimental data, or research conclusions. All technical content and scientific contributions presented in this work are the original work of the authors.